\documentclass[aps,pra,twocolumn,final,showpacs,superscriptaddress]{revtex4}
\usepackage[dvips]{graphicx}
\usepackage{amsmath}
\usepackage{float}
\usepackage{color}

\newcommand{\rmd}{{\rm d}}
\newcommand{\rme}{{\rm e}}
\newcommand{\rmi}{{\rm i}}
\newcommand{\mbfj}{{\mathbf{j}}}
\newcommand{\mbfp}{{\mathbf{p}}}
\newcommand{\mbfr}{{\mathbf{r}}}
\newcommand{\mbfT}{{\mathbf{T}}}

\newcommand{\Ai}{{\rm Ai}}
\newcommand{\Ci}{{\rm Ci}}
\newcommand{\Her}{{\rm H}}
\newcommand{\BesselJ}{{\rm J}}

\newcommand{\Qk}{{\rm Q}}
\newcommand{\Qi}{{\rm Qi}}
\newcommand{\PV}{{\rm P}}
\def\ket#1{|#1\rangle}

\begin{document}

\title{Quantum theory of an atom laser originating from a Bose-Einstein condensate or a Fermi gas in the presence of gravity}
\author{Tobias Kramer}
\affiliation{Instituto de F\'isica, Universidad Nacional Aut\'onoma de M\'exico, M\'exico D.F., M\'exico.}
\affiliation{Department of Physics, Harvard University, 17 Oxford Street, Cambridge, MA 02138, USA.}
\email{tobias.kramer@mytum.de}
\author{Mirta Rodr\'iguez}
\affiliation{Instituto de F\'isica, Universidad Nacional Aut\'onoma de M\'exico, M\'exico D.F., M\'exico.}
\affiliation{Clarendon Laboratory, University of Oxford, Parks Road, Oxford, OX1 3PU, United Kingdom.}
\email{m.pinilla1@physics.ox.ac.uk}
\date{May 8, 2006}
\begin{abstract}
We present a 3D quantum mechanical theory of radio-frequency
outcoupled atom lasers from trapped atomic gases in the presence of
the gravitational force. Predictions for the total outcoupling rate
as a function of the radio-frequency and for the beam wave function
are given. We establish a sum rule for the energy integrated
outcoupling, which leads to a separate determination of the coupling
strength between the atoms and the radiation field.

For a non-interacting Bose-Einstein condensate analytic solutions
are derived which are subsequently extended to include the effects
of atomic interactions. The interactions enhance interference
effects in the beam profile and modify the outcoupling rate of the
atom laser. We provide a complete quantum mechanical solution which
is in line with experimental findings and allows to determine the
validity of commonly used approximative methods.

We also extend the formalism to a fermionic atom laser and  analyze
the effect of superfluidity on the outcoupling of atoms.
\end{abstract}

\maketitle

\section{Introduction}

The possibility of creating an atom laser analogous to an optical
laser was considered immediately after the creation of atomic
Bose-Einstein condensates (BECs). Atom lasers can be operated by
continuously extracting small amounts of trapped atoms in a coherent
way. Atom lasers provide an important tool to analyze the properties
of trapped atoms and many experimental applications are expected due
to their coherence properties \cite{Choi2005a}. In principle they
offer the possibility to monitor the evolution of a BEC without the
need to switch off the trapping potential.

The first experimental realization of a BEC output coupler was
reported in Ref.~\cite{Mewes1997a}, where short radio-frequency (RF)
pulses changed the hyperfine state of the atoms. The inhomogeneous
magnetic trapping field separated the atoms into trapped and
outcoupled components. Using a series of RF pulses, a sequence of
coherent atom waves was formed.

A series of downward-falling output pulses analogous to a pulsed
laser was demonstrated in Ref.~\cite{Anderson1998a} using an optical
lattice. A BEC was loaded into a vertical standing wave created with
laser beams pointing in opposite directions. By lowering the depth
of the lattice, phase-coherent atoms from different wells tunneled
out of the traps and accelerated in the gravitational field.
Similarly, it is possible to release an extended wave packet from
a single optically trapped BEC by slowly lowering the trapping potential
\cite{Cennini2003a,Kramer2005a}.

A well-collimated quasi-continuous atom laser was achieved using a
stimulated Raman transition as outcoupling mechanism
\cite{Hagley1999a}. The two frequency Raman process imparts a
momentum kick to the extracted atoms allowing directional output
coupling. A sequence of overlapping matter wave packets were
extracted from a BEC using repeated Raman pulses.  Also the first
continuous high flux Raman atom laser has recently been reported
\cite{Robins2005a}.

Using an extremely stable novel magnetic trap, Bloch et al.
\cite{Bloch1999a} demonstrated a quasi-continuous output coupler for
magnetically trapped atoms. A weak RF field induces spin flips
between trapped and untrapped hyperfine states. The untrapped atoms
fall in the gravitational field producing a collimated atomic beam
whose duration is determined by the condensate size. The possibility
of continuous feeding the atomic source was demonstrated in
Ref.~\cite{Chikkatur2002a}. By using two different RFs on the same
condensate, the coherent spatial nature of the atom laser beam was
shown in Ref.~\cite{Bloch2000a}. The temporal coherence of
atom lasers was investigated in  Ref.~\cite{Koehl2001a}, and
more recently also the second order temporal correlation function
\cite{Oettl2005a}.

Since atom lasers generate coherent matter waves in the
gravitational field of the earth, an accurate description of the
quantum-mechanical propagation in a linear force field is an
important part of a theoretical model of an atom laser
\cite{Kramer2002a,Bracher2003a}. One dimensional models
\cite{Schneider1999a} are not sufficient for the characterization of
the beam wave function, which requires a fully three-dimensional
theory. Previous calculations of atom lasers rely mainly on
numerical integration without including interactions
\cite{Gerbier2001a}, or employ semiclassical approximations for the
propagation in the presence of gravity and a mean field potential
\cite{Busch2002a,Koehl2005a,Riou2005a}.

Recently, radial structures perpendicular to the gravitational field
have been observed in atom lasers \cite{Koehl2005a,Riou2005a}.
Similar structures have been predicted for smaller condensates
\cite{Kramer2002a,Kramer2003b,Kramer2003c}, where they are linked to
two-path interference in the presence of a linear force field.

In this paper, we formulate and apply a theory of an atom laser,
which employs the full 3D quantum mechanical propagator. After some
basic definitions in Sec.~\ref{sec:Definitions} we review and extend
in Sec.~\ref{sec:IdealLaser} the analytic solution for the beam
profile and the total current in the case of a non-interacting BEC.
The analytic solvable model of an atom laser supplied by an
``ideal'' BEC forms the basis for the inclusion of  mean field
potentials in Sec.~\ref{sec:GPLaser}. There, we derive a quantum
mechanical multiscattering theory for the interacting BEC.
We compare the quantum solution to experiments and approximative methods
like the Born series and the reflection approximation in order
to judge the underlying assumptions and shortcomings of
these approximations. The inclusion of a mean-field potential
changes the emission rate of an atom laser, but the rate still obeys
an important sum rule. The beam profile develops a radial
substructure, which can be obtained from our quantum
mechanical model and is in line with experimental observations.

The effect of spatially anisotropic traps is discussed in
Sec.~\ref{sec:GeometryEffects}, where we show how one-dimensional and
three-dimensional models are related.

In Sec.~\ref{sec:CurrHigh} we extend the formalism to the current
from higher modes in a harmonic trap. We apply it to a quasi
one-dimensional Fermi gas and discuss the effect
of fermionic superfluidity in both the current and the outcoupled
density profile.

The quantum source formalism provides a consistent framework for all
presented calculations. A quantum theory for the beam wave function
is of special importance for the coherent quantum control and
tailoring of matter waves \cite{Choi2005a}.

\section{The emission rate and the atomic beam wave function}\label{sec:Definitions}

The output coupling of magnetically trapped atoms can be understood
in terms of a spin flip of the magnetic hyperfine quantum number
$m_F$ \cite{Bloch1999a,Gerbier2001a} (for $^{87}$Rb atoms, the $F=1$
hyperfine level is commonly used). Initially, the atoms in the state
$|F=1,m_F=-1\rangle$ are in an eigenstate of the atomic trap
Hamiltonian in the presence of a static magnetic field $B_z$ and the
gravitational field $F=mg$ ($g\approx 9.81$ms$^{-2}$) along the
$z$-axis
\begin{eqnarray}\label{eq:Htrap}
H_{\rm trap}^{m_F=-1}
&=&\frac{\mbfp^2}{2m}+\frac{1}{2}m(\omega_x^2 x^2+\omega_y^2 y^2+\omega_z^2
z^2)\delta_{m_F,-1}\nonumber
\\
&&+m_F g_F \mu_B {\cal B}_z-Fz.
\end{eqnarray}
Here, $g_F$ denotes the Land\'e factor and $\mu_B$ the Bohr
magneton. The inhomogeneous magnetic field of the atom trap is
expanded in second order as a harmonic oscillator potential for the
$m_F=-1$ state. The gravitational field merely shifts the origin of
this oscillator along the $z$ direction and could be absorbed in the
quadratic term. The application of an additional oscillating
magnetic field with frequency $\nu$ and amplitude ${\cal B}'$ adds
to eq.~(\ref{eq:Htrap}) the time-dependent potential
$V(t)=-\boldsymbol{\mu}{\cal B}' \cos(\nu t)$ which causes
transitions of the spin to the magnetic quantum number $m_F=0$ (for
simplicity we will not consider the $m_F=1$ state). However, the
$|F=1,m_F=0\rangle$ state is no longer an eigenstate of the trapping
Hamiltonian, which only supports the $m_F=-1$ state. Instead its
evolution is governed by the Hamiltonian
\begin{eqnarray}\label{eq:Hcont}
H_{\rm grav}^{m_F=0}=\frac{\mbfp^2}{2m}-Fz,
\end{eqnarray}
which leads to the propagation away from the trap and the formation
of an atom laser beam. In the following we will assume a relatively
weak amplitude ${\cal B}'$ of the oscillating field, which makes it
possible to deplete the ground state of the trap over some time.
Since our main interest is the determination of the atom laser
wave function and the emission rate as a solution of the Schr{\"o}dinger equation,
we will treat the radio wave as a classical radiation field.

At this point we still have to solve for the time-dependent
eigenstates $\psi_{\rm trap}'$ and $\psi_{\rm grav}$ of a coupled
system, which have an energy difference of $\Delta E=E_{\rm
grav}-E_{\rm trap}$ and are coupled via $\gamma=\mu {\cal B}'$
\begin{eqnarray}
(\rmi\hbar\partial_t-H_{\rm grav})\psi_{\rm grav}(\mbfr,t)
 &=\gamma\;\rme^{-\rmi\Delta E t/\hbar}\psi_{\rm trap}'(\mbfr,t) \\
(\rmi\hbar\partial_t-H_{\rm trap})\psi_{\rm trap}'(\mbfr,t)
 &=\gamma\;\rme^{+\rmi\Delta E t/\hbar}\psi_{\rm grav}(\mbfr,t).
\end{eqnarray}
Here, we employed the rotating wave approximation. We split off the time dependence of the states
\begin{eqnarray}
\psi_{\rm grav}(\mbfr,t)&=\rme^{-\rmi E_{\rm grav}t/\hbar}\psi_{\rm
grav}(\mbfr)\\
\psi_{\rm trap}'(\mbfr,t)&=\rme^{-\rmi E_{\rm trap}t/\hbar}\psi_{\rm
trap}'(\mbfr),
\end{eqnarray}
in order to obtain the stationary equations
\begin{eqnarray}
(E_{\rm grav}-H_{\rm grav})\psi_{\rm grav}(\mbfr)
 &=\gamma\;\psi_{\rm trap}'(\mbfr) \label{eq:PsiStationaryCont}\\
(E_{\rm trap}-H_{\rm trap})\psi_{\rm trap}'(\mbfr)
 &=\gamma\;\psi_{\rm grav}(\mbfr).
\end{eqnarray}
Now we will use the assumption of a weak coupling in order to
replace the state $\psi_{\rm trap}'(\mbfr)$ by an eigenstate of
eq.~(\ref{eq:Htrap}) denoted by $\psi_0(\mbfr)$. Thus we break the
coupling between the two equations and are left with the evaluation
of the stationary Schr\"odinger equation
\begin{equation}\label{eq:SchroedingerSource}
(E_{\rm grav}-H_{\rm grav})\psi_{\rm grav}(\mbfr)=\gamma\;\psi_0(\mbfr)
\end{equation}
in the presence of an inhomogeneous source term
\begin{equation}
\sigma(\mbfr)=\gamma\;\psi_0(\mbfr).
\end{equation}
The time-independent source term is akin to the steady state
solution for the atom laser beam after the damping of transient
effects due to the initial switching on. In the following, we
restrict the discussion to the stationary case. The inhomogeneous
equation is readily solved by using the energy-dependent Green
function $G_{\rm grav}(\mbfr,\mbfr';E)$ for $H_{\rm grav}$. The
Green function is the solution of the Schr\"odinger equation for a
point inhomogeneity
\begin{equation}
(E-H_{\rm grav})G_{\rm grav}(\mbfr,\mbfr';E)=\delta(\mbfr-\mbfr'),
\end{equation}
and thus we obtain the wave function emitted from an extended source $\sigma(\mbfr)$ by a convolution integral:
\begin{equation}\label{eq:PsiContGreen}
\psi_{\rm grav}(\mbfr;E)=\int \rmd\mbfr'\,G_{\rm
grav}(\mbfr,\mbfr';E)\,\sigma(\mbfr').
\end{equation}
In the following we will suppress the subscript ${\rm grav}$, since
we are always interested in the properties of the atomic beam. We
take the energy $E$ as the difference between the energy of the
radiation field $h\nu$ minus the Zeeman splitting between the levels
$m_F=-1$ and $m_F=0$:
\begin{equation}\label{eq:energy}
E=h\nu-(E_{\rm grav}-E_{\rm trap}).
\end{equation}
Notice that due to continuous spectrum of $H_{\rm grav}^{m_F=0}$, an
output coupling is possible for a continuous range of energies.
However, for $E\rightarrow \pm \infty$ we will see that a sum rule
for the energy integrated outcoupling rate enforces a vanishing
outcoupling rate.
We now proceed to calculate the total current $J(E)$, which denotes the number of
atoms released per second. We do this by defining a current density
associated with the wave function (\ref{eq:PsiContGreen})
\begin{equation}\label{eq:CurrentDensity}
\mbfj(\mbfr;E)=
\frac{\hbar}{m}\Im\left[\psi(\mbfr;E)^{*}\,\nabla \psi(\mbfr;E)\right].
\end{equation}
By using eq.~(\ref{eq:SchroedingerSource}) it is straightforward to
derive the equation of continuity for this stationary problem:
\begin{equation}\label{eq:CurrentDivergence}
\boldsymbol{\nabla}\cdot\mbfj(\mbfr)=-\frac{2}{\hbar}\Im\{\sigma(\mbfr)^{*}\,
\psi(\mbfr)\}.
\end{equation}
As wanted, the inhomogeneity models a constantly emitting particle
source. The integration over a surface enclosing the source yields a
bilinear expression for the total probability current:
\begin{equation}\label{eq:CurrentTotal}
J(E)=-\frac{2}{\hbar}\Im
\left\{
\int\rmd\mbfr\,\int\rmd\mbfr'\,\sigma(\mbfr)^{*}\,G(\mbfr,\mbfr';E)\,
\sigma(\mbfr')
\right\}.
\end{equation}
Using the alternative representation of the Green function
\cite{Halperin1952a,Bracher2003a},
\begin{equation}
G(\mbfr,\mbfr';E)=\left\langle\mathbf{r}\left|
\PV\left(\frac{1}{E-H}\right)-\rmi\pi\delta(E-H)
\right|\mathbf{r}'\right\rangle.
\end{equation}
we can rewrite eq.~(\ref{eq:CurrentTotal}) as \cite{Heller1978a,Bracher2003a}:
\begin{equation}\label{eq:jtot_trace}
J(E)=\frac{2\pi}{\hbar}\langle\sigma|\delta(E-H)|\sigma\rangle
\end{equation}
The quantity $\langle \mathbf{r}|\delta(E-H)|\mathbf{r}\rangle$ is
the local density of states (LDOS) of the Hamiltonian $H$. For
initial states normalized to $\langle\psi_0|\psi_0\rangle = N$ an
important sum rule \cite{Kramer2002a} follows from
eq.~(\ref{eq:jtot_trace}):
\begin{equation}\label{eq:sumrule}
\int_{-\infty}^{\infty}\rmd E
J(E)=\frac{2\pi}{\hbar}\langle\sigma|\sigma\rangle =\frac{2\pi
\gamma^2 N}{\hbar}.
\end{equation}
One consequence of the sum rule is, that for any Hamiltonian we can
determine the interaction strength by simply summing up the total
current at different energies. In Ref.~\cite{Kramer2002a} this was used
to check the experimental reported coupling strength $\gamma$. Also
the finite value of the integral in eq.~(\ref{eq:sumrule}) restricts the output coupling to a
specific energy range. In the next sections, we derive analytic expressions for the Green function $G(\mbfr,\mbfr';E)$.

It is worth mentioning that the source formalism introduced here is
analogous to the first order perturbation theory in the coupling
Hamiltonian used for fermionic RF outcoupling in
\cite{Torma2000a,Bruun2001a,Kinnunen2004a}.

\section{The ideal atom laser}\label{sec:IdealLaser}

In this section we discuss an extension of the theory for an ideal
atom laser presented in Ref.~\cite{Kramer2002a}. The ideal case
forms the basis for the inclusion of interactions into the theory in
Sect.~\ref{sec:GPLaser}. For the Hamiltonian of the linear force
field (\ref{eq:Hcont}) there exists a closed analytic expression of
the Green function \cite{Dalidchik1976a,Li1990a,Gottlieb1991a}:
\begin{equation}\label{eq:GreenField}
G_{\rm grav}(\mbfr,\mbfr';E) =
\frac{m}{2\hbar^2}\frac{\Ci(u_+)\Ai'(u_-)-\Ci'(u_+)\Ai(u_-)}
{|\mbfr-\mbfr'|},
\end{equation}
where
\begin{eqnarray}
u_\pm&=&-\beta\left[2E+F(z+z')\pm F|\mbfr-\mbfr'|\right], \\
\beta     &=&\left[m/(4\hbar^2F^2)\right]^{1/3},
\end{eqnarray}
and ${\rm Ci}(x)={\rm Bi}(x)+\rmi\, {\rm Ai}(x)$. The application of
this Green function to an ideal atom laser from the ground state of
a BEC is discussed in detail in \cite{Kramer2002a} and extended to
include vortices in rotating BECs in \cite{Bracher2003a}. In
Sec.~4.2 of \cite{Kramer2002a}, an analytic solution for an atom
laser originating from a non-interacting isotropic BEC with $N$
atoms of Gaussian form
\begin{equation}\label{eq:BEC}
\sigma(\mathbf{r}) = \gamma \sqrt{N} a^{-3/2} \pi^{-3/4} {\rm e}^{-r^2/2a^2} \;,
\end{equation}
is derived. Using scaled variables
\begin{eqnarray}\label{eq:ScaledVariables}
(\xi,\upsilon,\zeta)=\beta F (x,y,z), \quad
\epsilon=-2\beta E,\quad
\alpha=\beta F a,\\
\tilde{\zeta}=\zeta+2\alpha^4, \quad
\tilde{\rho}^2  =\xi^2+\upsilon^2+\tilde{\zeta}^2, \quad
\tilde{\epsilon}=\epsilon+4\alpha^4,
\end{eqnarray}
and the special functions defined in \cite{Bracher2003a}, App.~B
\begin{eqnarray}
\Qk_1(\rho,\zeta;\epsilon)&=&-\frac{1}{2\rho}
\big[\Ci(\epsilon-\zeta-\rho)\Ai'(\epsilon-\zeta+\rho)\nonumber\\
&&-\Ai(\epsilon-\zeta+\rho)\Ci'(\epsilon-\zeta-\rho)\big]\\
\Qk_1^{\rm near}(\rho,\zeta;\epsilon)&=&-\int_0^{2\alpha^2}\rmd u
\frac{\rme^{-\frac{\rho^2}{u}-u(\epsilon-\zeta)+\frac{u^3}{12}}}{2\pi^{3/2}u^{
3/2}}\\
\Qi_1(\epsilon)&=&{[\Ai'(\epsilon)]}^2-\epsilon{[\Ai(\epsilon)]}^2,
\end{eqnarray}
one obtains analytical expressions for the total current
(\cite{Bracher2003a}, eq.~(73))
\begin{equation}\label{eq:jtot_ideal}
J_{\rm ideal}(E) = \frac{8}{\hbar}\, \beta{(\beta F)}^3
\Lambda(\tilde{\epsilon})^2 \Qi_1(\tilde\epsilon),
\end{equation}
and the beam wave function
\begin{equation}\label{eq:psi_ideal}
\psi_{\rm ideal}(\mbfr;E) = -4\beta{(\beta F)}^3 \Lambda(\tilde{\epsilon})
\big[\Qk_1(\tilde\rho, \tilde\zeta; \tilde\epsilon)+\Qk_1^{\rm near}(\tilde\rho,
\tilde\zeta;
\tilde\epsilon)\big].
\end{equation}
\begin{figure}[t]
\includegraphics[width=\columnwidth]{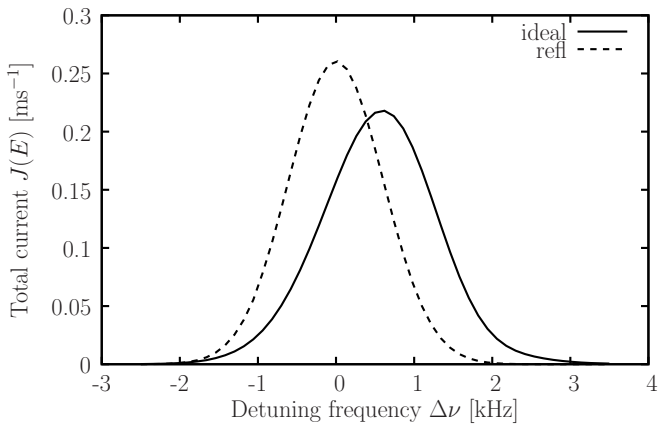}
\caption{Total current (per atom) as a function of radio frequency
detuning $\Delta\nu=E/h$ from eq.~(\ref{eq:jtot_ideal})
(for the sign of the detuning see the
endnote \cite{EndNoteDetuningSign}). Also shown is
the reflection approximation (dashed line).
The area underneath the curves is fixed by the sum rule (\ref{eq:sumrule}).
Parameter:
$a=0.4\;\mu$m,
$\gamma/h=100$~Hz,
$F=m_{\text{Rb}}\,g$,
with $g=9.81$~m/s$^2$, and
$m_{\text{Rb}}=87$~u.\label{fig:jtot_gauss}}
\end{figure}
In addition to eq.~(72) in Ref.~\cite{Bracher2003a}, we explicitly
add the near field contribution denoted by $\Qk_1^{\rm near}$. For the
non-interacting case, this term has no influence on the beam profile
outside the condensate region, whereas  we have to include it for
the interacting model in the next section. In both expressions, the
source strength $\Lambda(\tilde\epsilon)$ is strongly energy- and
size-dependent:
\begin{equation}
\label{eq:Atom4}
\Lambda(\tilde\epsilon) = \sqrt N\, \gamma {(2\sqrt{\pi}a)}^{3/2} {\rm
e}^{2\alpha^2 \left(\tilde\epsilon -
4\alpha^4/3 \right)} \;.
\end{equation}

\begin{figure}[t]
\includegraphics[width=0.49\columnwidth]{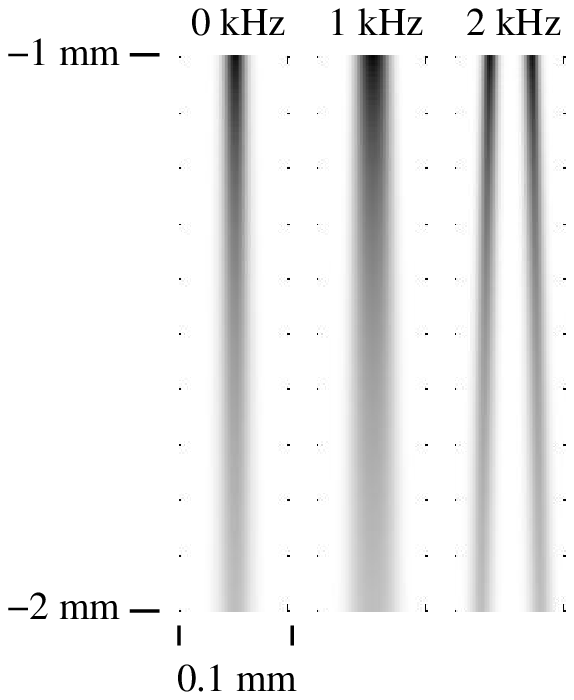}
\includegraphics[width=0.49\columnwidth]{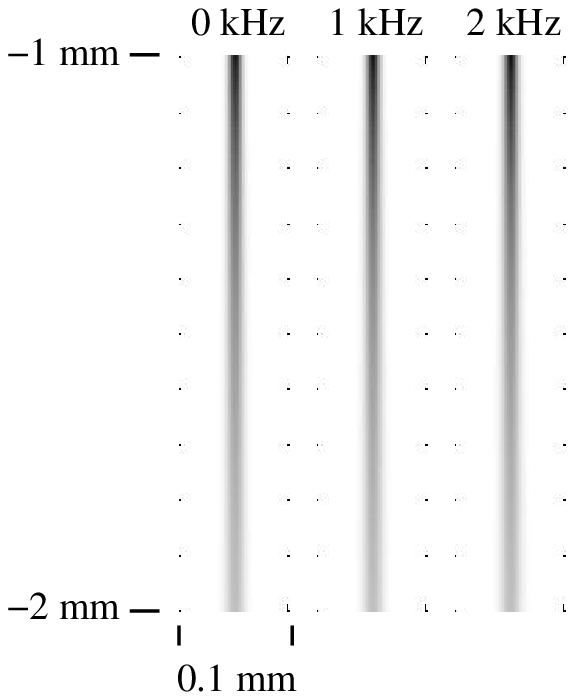}
\caption{Atom laser beam profile at different
detuning energies $\Delta\nu=E/h$ from eq.~(\ref{eq:psi_ideal})
for a relatively small BEC (left panel, $a=0.4\;\mu$m) and a
larger BEC (right panel, $a=0.8\;\mu$m).
For the sign of the detuning see the
endnote \cite{EndNoteDetuningSign}. Shown is a cut through the middle
of the beam along the vertical axis from $1$~mm to $2$~mm below the BEC
for the detuning frequencies $(0,1,2)$~kHz. The beam profile widens and
develops a transverse substructure for larger energies. There is
rotational symmetry about the vertical-middle axis of each profile.
Parameter:
$a=0.4\;\mu$m, $F=m_{\text{Rb}}\,g$, with $g=9.81$~m/s$^2$, and
$m_{\text{Rb}}=87$~u.\label{fig:jz_gauss}}
\end{figure}

Let us briefly summarize the main features of the ideal atom laser model (see also Section~\ref{sec:sc_atomlaser}):
\begin{itemize}
\item As shown in \cite{Kramer2002a}, Sec.~3.3, the beam wave function can be mapped back to a virtual tunneling point source, which is closely related to the Green function.
\item For small condensates (radius $a<\frac{m F a^4}{2\hbar^2}$ in the direction of the gravitational field, i.e.\ about  $0.5\;\mu$m  for a Rb BEC), additional modulations in the total current appear and the beam wave function develops an interference structure. The emerging pattern can be explained in terms of two-path interference in the linear force field. A typical sequence of beam profiles for different detuning energies is shown in left panel of Fig.~\ref{fig:jz_gauss}.
\item For large condensates (radius $a>\frac{m F a^4}{2\hbar^2}$ in the direction of the gravitational field), the energy dependency of the total current reflects the density distribution of the source.
The beam wave function is featureless over the whole energy range and well described by an Gaussian profile \cite{Kramer2002a}, right panel of Fig.~\ref{fig:jz_gauss}.
\end{itemize}
We are not aware of the experimental observations of atom laser
beams from small condensates, although high-quality magnetic microtraps
are able to produce the required BECs \cite{Schumm2005a}.

For larger condensates one can derive
approximative expressions for the total current. Of special
simplicity is the so-called reflection approximation, which was
developed in the theory of Franck-Condon factors
\cite{Herzberg1950a}. It is analogous to the local density
approximation commonly used in BEC theory and consists of neglecting
the kinetic energy term in the Hamiltonian (\ref{eq:Hcont}).
Now  eq.~(\ref{eq:jtot_trace}) simplifies to
\begin{eqnarray}\label{eq:jtot_ideal_refl}
J_{\rm ideal}^{\rm refl}(E)
&=&\frac{2\pi}{\hbar} \langle\sigma|\delta(E+Fz)|\sigma\rangle\nonumber\\
&=&\frac{2\pi}{\hbar} \int \rmd\mbfr {|\sigma(\mbfr)|}^2 \delta(E+Fz).
\end{eqnarray}
The reflection approximation states that the current is proportional to a slice through the initial density distribution along a plane height $z=E/F$. It is possible to justify this approximation as a limit of the quantum solution (see \cite{Kramer2002a}, eq.~(40)). In principle one can also calculate quantum corrections to eq.~(\ref{eq:jtot_ideal_refl}) (see \cite{Heller1978a,Huepper1998a,Japha2002a,Kramer2003b}), but the resulting (asymptotic) series can diverge for a finite number of terms (see Ref.~\cite{Huepper1998a}, Sec.~VI).

The range of validity is limited by the requirement for a large
spatial overlap between the initial wave function and the outgoing
wave function in order to average over the oscillations of the Airy
function in Eq.~(\ref{eq:GreenField}). If the width of the initial
wave function is smaller than the first oscillation period of the
Airy function (given by approx.\ $1/(\beta F)$), the oscillations in
the outgoing wave function carry over to the total current (see
Sect.~\ref{sec:GeometryEffects}).

In Fig.~\ref{fig:jtot_gauss} we compare the reflection approximation with the quantum mechanical result eq.~(\ref{eq:jtot_ideal}).

While the reflection approximation can in certain limits reproduce the total current distribution, it cannot yield information about the atomic beam profile. The slicing picture may be suggestive for the idea that atoms are only outcoupled at the slice given by the condition $z=E/F$. However, this assumption is not supported by the quantum mechanical result (\ref{eq:psi_ideal}). Contrary to the picture of atoms leaving the condensate with zero momentum along a slice (which would in a (semi-)classical picture imply that the radial profile of the beam at all distances from the BEC is identical to the density profile of the BEC), the beam profile spreads out as shown in Fig.~\ref{fig:jz_gauss}. The summation over the infinitely many starting points distributed over the complete BEC is represented by a single point above the center of the condensate, and not by a planar surface at $z=E/F$.
The semiclassical picture is further discussed in Sec.~\ref{sec:sc_atomlaser}.

\subsection*{Simultaneous outcoupling with two different radio frequencies.}

\begin{figure}[t]
\includegraphics[width=0.7\columnwidth]{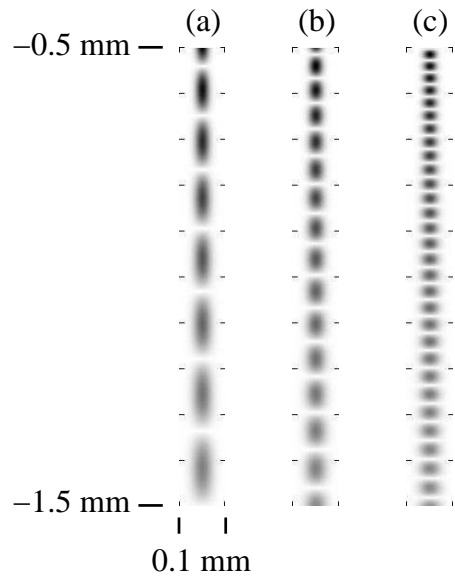}
\caption{Squared beam wave function for simultaneous output coupling with
two different radio frequencies. Shown is the density given by
eq.~(\ref{eq:twobeams}). The outcoupling frequencies
$\Delta\nu_{1,2}$ are (a) $\pm 0.5$~kHz, (b) $\pm 1.0$~kHz, and (c)
$\pm 2.0$~kHz. The number of longitudinal interference  fringes is
proportional to the difference in the detuning frequencies.
Parameter: $a=0.8\;\mu$m, $F=m_{\text{Rb}}\,g$, with
$g=9.81$~m/s$^2$, and
$m_{\text{Rb}}=87$~u.\label{fig:quantumfaucet}}
\end{figure}

Experiments which outcouple atomlasers from a BEC with two different radio frequencies at the same time \cite{Bloch2000a} have shown the appearance of longitudinal interference structures.  Using the quantum mechanical theory of the previous section, the resulting atom laser beam is described by the coherent superposition of two stationary beams with different energy originating from the same virtual point source:
\begin{eqnarray}\label{eq:twobeams}
{|\psi_{\rm ideal}^{\rm two~RF}(\mbfr,t;\nu_1,\nu_2)|}^2&=&
\big| \psi_{\rm ideal}(\mbfr;h\nu_1)e^{\rmi h \nu_1 t}\\
&&+ \psi_{\rm ideal}(\mbfr;h\nu_2)e^{\rmi h \nu_2 t} \big|^2.\nonumber
\end{eqnarray}
The superposition of the two beam wave functions leads to a time-dependent oscillation of the density profile, which reproduces the observed longitudinal interference structure as shown in Fig~\ref{fig:quantumfaucet}. The use of multiple radio frequencies provides an important tool for tailoring the atomic beam wave function.

\section{Mean field effects in the atom laser}\label{sec:GPLaser}

In this section we extend the quantum theory to include interactions
between the BEC and the emitted atom laser beam (interactions within
the atomic beam can be neglected since the density is much smaller
than inside the BEC). A commonly used approach to include
interactions in the description of a BEC is the addition of the
(repulsive) mean-field potential via the Gross-Pitaevskii equation
\cite{Dalfovo1999a}:
\begin{equation}
V_{\rm GP}(\mbfr)=g_{\rm sc}{|\psi_0(\mbfr)|}^2,
\end{equation}
where $g_{\rm sc}$ denotes the interaction strength and is related
to the scattering length $a_{sc}$ and the number of atoms in the BEC
$N$ via
\begin{equation}
g_{\rm sc}=4\pi\hbar^2 N a_{\rm sc}/m.
\end{equation}
In principle the density dependent term leads to a non-linear
Schr{\"o}dinger equation. However, for the theory of the atom laser
we will treat the BEC (and ${|\psi_0(\mbfr)|}^2$) as unchanged
during the outcoupling process. Therefore we just have to modify the
Hamiltonian for the propagating state by the mean field potential of
the BEC:
\begin{equation}\label{eq:hgrav+gp}
H_{\rm GP}=\frac{\mathbf{p}^2}{2m}-Fz+g_{\rm sc}{|\psi_0(\mbfr)|}^2.
\end{equation}
The additional repulsion will lead to a broadening of the beam
compared to the non-interacting case, as we will show next. The
repulsive interaction also leads to a change in the BEC density
distribution itself. In the following we will retain the Gaussian
form of the condensate, but the half-width $a$ of the BEC should be
viewed as a parameter, which can be obtained i.e.\ by minimizing the
Gross-Pitaevskii energy-functional \cite{Baym1996a}. For large
condensates, a Thomas-Fermi like profile of the density is more
appropriate than the Gaussian approximation, whereas for condensates
in strongly confining traps a Gaussian profile is a fairly good
approximation \cite{Schumm2005a}, because larger trapping
frequencies increase the ratio of the kinetic energy vs.\ the
interaction energy for a constant maximum condensate density (see
Ref.~\cite{Pethick2002a}, Sect.~6.2). However, the following
discussion and comparison of different methods for the atom laser
rate and profile is in principle not limited to an initially
Gaussian density distribution.

\subsection{Quantum Theory}

The Green function of the Hamiltonian (\ref{eq:hgrav+gp}) is not
available in analytic form. In principle, the Born series could be
used to construct the Green function:
\begin{eqnarray}
G_{\rm GP}&=&G_{\rm grav}+G_{\rm grav}V_{\rm GP}G_{\rm grav}\nonumber\\
&&\quad+G_{\rm grav}V_{\rm GP}G_{\rm grav}V_{\rm GP}G_{\rm grav}+\ldots
\end{eqnarray}
As we will see below, for a typical BEC this series converges very
slowly. An alternative approach consists in decomposing the mean
field potential in terms of a $\delta$-lattice
\begin{equation}
V_{\delta}(\mbfr)=\sum_{j=1}^N V_{\rm GP}(\mbfr_j)
\Delta\mbfr\;\delta(\mbfr-\mbfr_j),
\end{equation}
where $\Delta\mbfr$ denotes the volume element of each lattice site.
In order to mimic a continuous potential, the lattice spacing must be smaller than
the typical oscillation length of the Green function, which is given by
$\lambda\approx 1/(\beta F)$. Numerically, convergence has been checked
by a set of calculations with subsequently reduced lattice spacing.
For our calculations, we used a spacing of $0.15 \mu$m, which is about $\lambda/4$.
The $\delta$-lattice is algebraically solvable via the transition
(T-) matrix method (for a compact derivation see
Ref.~\cite{Donner2004a}, App.~D). The T-matrix involves only the
known Green function of the linear potential
\begin{equation}
\mbfT(E)^{-1}=
\left\{
\begin{array}{ll}
-G_{\rm grav}(\mbfr_j,\mbfr_k;E)                                   & j\neq k,\\
{\left[ V_{\rm GP}(\mbfr_j)\Delta\mbfr \right] }^{-1}-G_{\rm grav}^{\rm
norm}(\mbfr_j,E)& j=k
\end{array}
\right.
\end{equation}
and the renormalized Green function $G_{\rm grav}^{\rm norm}(\mbfr;E)$ for
$\mbfr=\mbfr'$ (\cite{Donner2004a}, (D21)):
\begin{equation}
G_{\rm grav}^{\rm norm}(\mbfr,E)=\frac{m\beta F}{\hbar^2}
\left[u\Ci(u)\Ai(u)-\Ci'(u)\Ai'(u)\right],
\end{equation}
with $u=-2\beta(E+Fz)$. The resulting Green function reads
\begin{eqnarray}
G_{\rm GP}(\mbfr,\mbfr';E)&=&G_{\rm grav}(\mbfr,\mbfr';E)\\\nonumber
&&\hspace{-2cm}
+\sum_{j,k=1}^{N}G_{\rm grav}(\mbfr,\mbfr_j;E) \mbfT_{jk}(E,\mbfr_j,\mbfr_k)
G_{\rm grav}(\mbfr_k,\mbfr';E).
\end{eqnarray}
Using the Green function $G_{\rm GP}$, we proceed to calculate the total current
from eq.~(\ref{eq:CurrentTotal}):
\begin{eqnarray}\label{eq:jtot_gp}
J_{\rm GP}(E)
&=&-\frac{2}{\hbar}\Im\big[\langle\sigma| G_{\rm grav} |
\sigma\rangle\nonumber\\
&&
+\sum_{j,k}\mbfT_{jk}(E)
\psi_{\rm ideal}(\mbfr_j;E)\psi_{\rm ideal}(\mbfr_k;E)\big]\nonumber\\
&=&J_{\rm ideal}(E)+J_{\rm GP}^{\delta}(E).
\end{eqnarray}
The sum rule (\ref{eq:sumrule}) enforces that the changes of the current
vanish upon integration over the outcoupling frequency:
\begin{equation}
\int_{-\infty}^\infty\rmd E\,J_{\rm GP}^{\delta}(E)=0.
\end{equation}
Similarly, the beam wave function (see eq.~(\ref{eq:PsiContGreen}))
is given by the sum of the ideal profile and an interaction term
\begin{eqnarray}\label{eq:psi_gp}
\psi_{\rm GP}(\mbfr;E)
&=&\psi_{\rm ideal}(\mbfr;E)\nonumber\\
&&
+\sum_{j,k}\mbfT_{jk}(E)
G_{\rm grav}(\mbfr,\mbfr_j;E)\psi_{\rm ideal}(\mbfr_k;E)\big]\nonumber\\
&=&\psi_{\rm ideal}(\mbfr;E)+\psi_{\rm GP}^{\delta}(\mbfr;E).
\end{eqnarray}

\subsection{First order Born and reflection approximation}

\begin{figure}[t]
\includegraphics[width=\columnwidth]{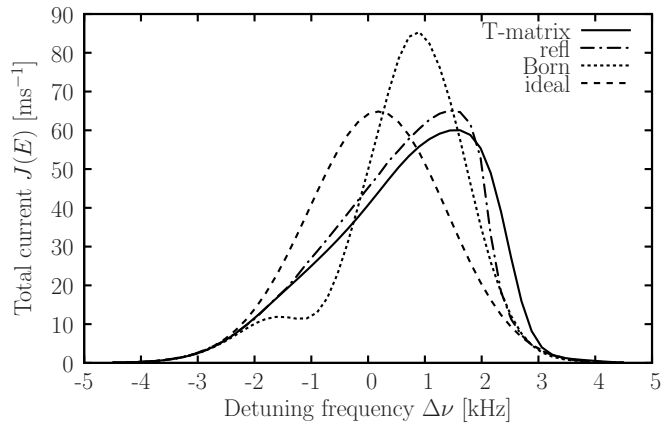}
\caption{Total current for a BEC of $N=500$ atoms as a function of
the detuning frequency $\Delta\nu=E/(2\pi\hbar)$. We compare the
results using the T-matrix method (\ref{eq:jtot_gp}) (with
1309 lattice points spaced at distances of $0.15 \mu$m),
the reflection approximation
(\ref{eq:jtot_gp_refl}), the first order Born approximation (\ref{eq:tmatrix_gp_born}),
and the non-interacting result
(\ref{eq:jtot_ideal}). Parameter:
$N=500$ atoms,
$a_{\rm scat}=5.77$~nm,
$a=0.8~\mu$m,
$\gamma/h=100$~Hz,
$F=m_{\text{Rb}}\,g$, with
$g=9.81$~m/s$^2$,
and $m_{\text{Rb}}=87$~u.\label{fig:jtot_gp}}
\end{figure}

Within the T-matrix approach, we obtain the first order Born approximation by setting
\begin{equation}\label{eq:tmatrix_gp_born}
\mathbf{T}_{jk}^{\rm Born}(E)=\delta_{jk} V_{\rm GP}(\mbfr_j) \Delta\mbfr.
\end{equation}
In general, the first order approximation is not sufficient for an accurate description,
as shown in Fig.~\ref{fig:jtot_gp}.

In analogy to the non-interacting case, we can include the mean
field potential in the reflection approximation. The previously planar
slices are now distorted, depending on the density of the condensate wave function:
\begin{equation}\label{eq:jtot_gp_refl}
J_{\rm GP}^{\rm refl}(E)=\frac{2\pi}{\hbar} \int \rmd\mbfr {|\sigma(\mbfr)|}^2
\delta(E+F z-V_{\rm GP}(\mbfr)).
\end{equation}

\subsection{Comparison of the quantum theory and approximative methods}

In Fig.~\ref{fig:jtot_gp} we compare the results of the different
methods for the total current as a function of the detuning
frequency. We choose a Gaussian condensate with half width
$a=0.8~\mu$m, for which we expect a tunneling behavior up to
detuning energies of $F z_0\approx8~$kHz. The ideal
(non-interacting) total current reflects the Gaussian density
profile of the condensate and attains therefore a symmetric shape.
The inclusion of the mean-field potential shifts the maximum of the
total current to higher energies, as shown by the quantum mechanical
T-matrix calculation. The reflection approximation works
surprisingly well, whereas the first order Born approximation gives
a misleading result with an additional hump. Notice that the area
underneath all curves is the same, as required by the sum
rule~(\ref{eq:sumrule}). The shift to higher values of the detuning
frequency in the maximum of the output coupling is in agreement with
experimental results reported in \cite{Bloch1999a,Gerbier2001a}
(note that the definition of the sign of the detuning frequency used
here is opposite from the one used in the experimental work).

\begin{figure}[t]
\begin{center}
\includegraphics[width=0.49\columnwidth]{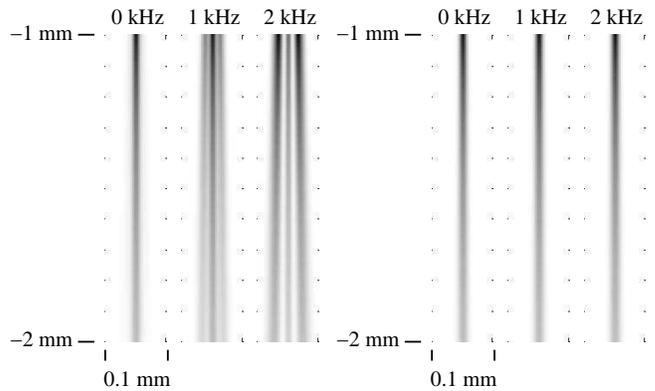}
\includegraphics[width=0.49\columnwidth]{fig_2b}
\end{center}
\caption{
Left panel: Atom laser beam profile from Eq.~(\ref{eq:psi_gp}) at different
detuning energies $\Delta\nu=E/h$ including interactions ($N=500$ atoms, $a_{\rm scat}=5.77$~nm),
other parameters as in Fig.~\ref{fig:jtot_gp}. The beam profile broadens and
develops a transverse substructure for larger energies.
Right panel:
Atom laser profiles for the same detuning frequencies
and condensate width, but without interactions.
The transverse interference pattern is not present.
For the sign of the detuning see the
endnote \cite{EndNoteDetuningSign}. Shown is a cut through the middle
of the beam along the vertical axis from $1$~mm to $2$~mm below the BEC
for the detuning frequencies $(0,1,2)$~kHz.
There is rotational symmetry about the vertical-middle axis of each profile.
\label{fig:jz_gp}
}
\end{figure}

Due to the effective negative initial kinetic energy, the beam
profile of the non-interacting atom laser has a Gaussian profile,
without a radial substructure (see \cite{Kramer2002a}). However, the
presence of the repulsive mean field potential affects the beam
profile considerably as shown in Fig.~\ref{fig:jz_gp}. In the
transverse direction a substructure develops, which has been
observed experimentally \cite{Koehl2005a,Riou2005a}. In a simple
one-dimensional picture, the widening of the beam profile as
compared to the non-interacting case has been attributed to the
repulsive hump in the potential acting as a diverging lens
\cite{Busch2002a}. The three-dimensional quantum mechanical picture
is more involved, since the T-matrix approach includes multiple
scattering events and no simple semiclassical interpretation in
terms of trajectories is available.

\subsection{Semiclassical models for the beam profile}\label{sec:sc_atomlaser}

The appearance of an interference structure in matter wave
experiments can be linked to the possibility of multiple paths from
the source (or emitter) of the wave to the location of the detector.
In the presence of a linear force field, the classical double slit
experiment for electrons \cite{Moellenstedt1959a} was carried out
by Blondel et al \cite{Blondel1996a,Blondel1999a} without actually
constructing a material double slit. The uniform field environment
provides (for positive initial kinetic energy) a region, in which
there are two paths connecting the source with the target
\cite{Galilei1638a}. In the semiclassical approximation of the energy
Green function all classical allowed paths carry a complex amplitude
(determined by the classical action) and are added coherently
\cite{Berry1972a,Demkov1982a,Bracher1998a,Bracher2003a}.

In contrast to the classical analysis of the trajectories from a single point in space to another
point, a spatially extended source region, like a BEC, seems to require the
addition of infinitely many paths leading from every point of the
source region to the target point. Remarkably, the single point
interference pattern is not destroyed by this averaging process.
The reason is that, similar to the technique of virtual point sources in optics,
one may replace the extended BEC by a single point source which is located at a distance
\begin{equation}\label{eq:shift_z0}
z_0=-\frac{m F a^4}{2\hbar^2}=-\frac{g}{2\omega^2}
\end{equation}
above the center of the BEC \cite{Kramer2002a} (we used
$a=\sqrt{\hbar/(m\omega)}$). The focal point of the parabola given
by the trapping potential (converted to spatial units) $V_{\rm
trap}(z)/F=\frac{1}{2 F} m\omega^2 z^2$ coincides with the location
of the  virtual point source at a distance $z_0$ from the
condensate. Due to the shift upwards in the gravitational field, the
initially available kinetic energy is reduced by the potential
energy at the shifted location
\begin{equation}
E_{\rm kin}=h\Delta\nu-V_{\rm GP}(0,0,z_0)-|F z_0|.
\end{equation}
Here, we also included the potential term due to the mean field of the
condensate atoms, which creates a hump in the otherwise planar
potential surface of the linear gravitational field. For a large
condensate, the initial shift $|z_0|\gg a$ leads to a virtual point
source which is actually located several half widths $a$ apart from
the center of the BEC. The resulting initial kinetic energy is
negative for detuning energies in the range of energies for which we
expect a significant total current $h\Delta\nu < F a$. For the
semiclassical analysis we note that no classical trajectories exist
up to the turning surface, which is given by the implicit equation
\begin{equation}\label{eq:CausticSurface}
h\Delta\nu-V_{\rm GP}(0,0,z)-|F z|=0.
\end{equation}
In principle, one can study the classical trajectories which start
from this caustic surface and end at a given target point.
The possibility of multiple trajectories leads to a coherent
sum over the corresponding classical actions \cite{Busch2002a,Riou2005a}.
However, a caustic surface is not an ideal starting point for a semiclassical
analysis, since the specification of an initial position and
simultaneously a definite momentum $\mbfp=0$ is not compatible with
the uncertainty relation \cite{Bracher1998a}. The unknown initial
phase and weight of the manifold of classical trajectories starting
from the caustic surface presents another difficulty for a well
defined semiclassical description.

\section{Geometry effects in the atom laser}\label{sec:GeometryEffects}

So far we have only considered spherically symmetric clouds of
atoms. In principle one can tune the magnetic trapping potential and
thus vary the frequencies of the harmonic trap. Experimentally it is
possible to obtain quasi one-dimensional (1D) systems
\cite{Goerlitz2001a,Schreck2001a} by tuning one of the frequencies
of the three-dimensional (3D) trap to a value which is much smaller
than the other frequencies or by using optical lattices to create
arrays of smaller 1D systems \cite{Greiner2001a}. A 1D gas with the
long axis in the direction of the gravitational field is
characterized by $\omega_z \ll
\omega_{\perp}=(\omega_x^2+\omega_y^2)^{1/2}$. The propagation
occurs in the three dimensional space and thus it is not sufficient
to consider merely the propagation in the gravitational field along
the beam axis. While the total energy is conserved, it falls into
parts related to propagation in the direction of the gravitational
field and in the perpendicular plane respectively. In the following
we analyze the relationship between three-dimensional and
one-dimensional calculations. For simplicity we do not include mean
field effects in the calculations.

We consider the current and outcoupled wave function for an initial state of the form
\begin{equation}\label{eq:iniz0}
\psi_{\rm ini}(x,y,z)=\langle \mbfr |\psi_{\rm ini} \rangle
=\psi^{\perp}_0(x,y)\psi_0(z),
\end{equation}
where $\psi_0(z)$ and $\psi^{\perp}_0(x,y)=1/\sqrt{\pi
a_{\perp}^2}e^{-(x^2+y^2)/{2a_{\perp}^2}}$ are the ground states of
the harmonic oscillator with half width $a=\sqrt{\hbar/(m
\omega_z)}$ and $a_{\perp}=\sqrt{\hbar/(m \omega_{\perp})}$.
The connection between the 1D and the 3D Green function of the
linear gravitational field is given by a convolution integral over
the transverse momenta $k$ and $k'$
\begin{eqnarray}\label{eq:G3d1d}
G^{3D}_{\rm grav}(\mbfr,\mbfr';E) &=& \frac{1}{{(2\pi)^2}}\iint\rmd
k\rmd k'\;\rme^{-\rmi k(x-x')-\rmi
k'(y-y')}\times\nonumber \\
&& G_{\rm grav}^{1D}\left(z,z';E-\frac{\hbar^2(k^2+k^{\prime
2})}{2m}\right),
\end{eqnarray}
where (\cite{Donner2004a}, eq.~(B2))
\begin{equation}
G_{\rm grav}^{1D}(z,z';E)=-4\pi \beta^2 F \Ci(u_{+})\Ai(u_{-}).
\end{equation}
For the total current (\ref{eq:CurrentTotal}) we evaluate the
expectation value of the Green function with respect to the initial
state $\psi_{\rm ini}$. The resulting convolution integral reads
\begin{equation}\label{eq:j_3d}
J^{3D}(E) = 2 a_\perp^2\int_0^\infty\rmd
k_\perp\,k_\perp\rme^{-a_\perp^2 k_\perp^2}
J^{1D}\left(E-\frac{\hbar^2 k_\perp^2}{2m}\right),
\end{equation}
where
\begin{equation}\label{eq:j_1d}
J^{1D}\left(E\right)=\frac{16 \beta \pi^{3/2}\alpha\gamma^2
N}{\hbar}e^{-\frac{16}{3}\alpha^6+4\alpha^2
\tilde{\epsilon}}\Ai(\tilde{\epsilon})^2
\end{equation}
is derived in eq.~(\ref{eq:CurrentTotal1Dz}).\\
\begin{figure}[t]
\includegraphics[width=\columnwidth]{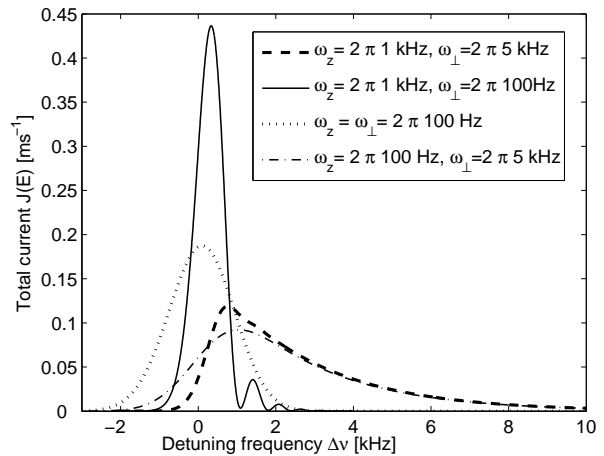}
\caption{Total current (per atom) as a function of the detuning frequency
$\Delta\nu=E/h$. Shown is the 3D total current (\ref{eq:j_3d}) for
several geometries.
Parameters: $\gamma/h = 100$~Hz,
$m_{\text{Na}}=23$~u.\label{fig:jtot_3d1d}}
\end{figure}
In Fig.~\ref{fig:jtot_3d1d} we show the 3D total currents for a big
condensate ($|z_0|>a$ or $Fa > 2 \hbar\omega_z$, see
eq.~(\ref{eq:shift_z0})) whose width along the gravitational axis is
$a=2.1~\mu$m ($\omega_z=2 \pi 100$Hz) and for a small BEC with
$a=0.66~\mu$m ($\omega_z=2 \pi 1$kHz). Fig.~\ref{fig:jtot_3d1d}
shows the current for Na atoms instead of Rb, since then the
condition for small condensates is fulfilled for smaller trapping
frequencies.
In the reflection approximation eq.~(\ref{eq:jtot_ideal_refl}) we
expect that the outcoupling window is just determined by the
condensate extension along the gravitational field and not changed
for the one-dimensional and three-dimensional case.

However, the convolution integral (\ref{eq:j_3d}) of the 1D current
with an exponentially decaying function whose width is proportional
to $\omega_{\perp}$ predicts in general a different form of the 1D
and 3D current. This effect can be clearly seen in
Fig.~\ref{fig:jtot_3d1d}. Only for a large condensate the 3D current
retains a Gaussian shape (see the dotted line in
Fig.~\ref{fig:jtot_3d1d}). A tight confinement in the transverse
direction results in a spread of the total current as shown by the
dot-dashed line in Fig. \ref{fig:jtot_3d1d}.

For small condensates the situation changes completely. Now the
total current in 1D is modulated by the zeroes of the Airy function
in eq.~(\ref{eq:j_1d}). The convolution with the transverse momentum
tends to wash out the zeroes if $\omega_{\perp}$ is big enough
(dashed line in Fig.~\ref{fig:jtot_3d1d}), but there might still
exist regions of almost complete suppression of the total current
within the outcoupling energy window (see the solid line in
Fig.~\ref{fig:jtot_3d1d}).

One can also separate the expressions for the 1D and 3D outcoupled
wavefunctions. The outgoing wave function perpendicular to the
gravitational force is given by
\begin{eqnarray}
\psi_{\rm out}^\perp(x,y)&=& \frac{1}{4a_\perp\pi^{5/2}} \iint\rmd
k\rmd k'
\iint\rmd x'\rmd y'\; \times\nonumber \\
&&\rme^{-\rmi k(x-x')-\rmi k'(y-y')-\frac{x^{\prime 2}+y^{\prime 2}}{2a_\perp^2}}\\
&=&\int\rmd k_\perp\frac{a_\perp k_\perp }{\sqrt{\pi}}\rme^{-\frac{a^2 k_\perp^2}{2}} 
\BesselJ_0(k_\perp\sqrt{x^2+y^2}),\nonumber
\end{eqnarray}
where $\BesselJ_0$ denotes the Bessel function of $0th$ order. The corresponding 3D wave-function reads:
\begin{eqnarray}\label{eq:psi3d}
\psi_{\rm out}^{3D}(\mbfr;E)
&=&\frac{a_\perp}{\sqrt{\pi}}\int_0^\infty\rmd k_\perp\,
k_\perp\rme^{-\frac{a_\perp^2 k_\perp^2}{2}}
\BesselJ_0(k_\perp\sqrt{x^2+y^2})
\times\nonumber\\
&&\psi_{\rm out}^{1D}(z;E-\frac{\hbar^2 k_\perp^2}{2m})
\end{eqnarray}
where
\begin{eqnarray}
\psi_{\rm out}^{1D}(z;E)&=&4\sqrt{2 F\alpha N} \gamma
\beta^{3/2}\pi^{5/4}e^{-\frac{8}{3}\alpha^6+2\alpha^2
\tilde{\epsilon}}\times\nonumber\\
&&\quad\Ai(\tilde{\epsilon})\Ci(\tilde{\epsilon}-2\tilde{\xi}).\label{eq:f_1D}
\end{eqnarray}
Similar to the total current, the three-dimensional wavefunction is given by the
convolution of the one-dimensional wavefunction and a exponentially decaying
function whose width is proportional to $\omega_{\perp}$.

\section{Current from higher modes in a harmonic trap. Fermionic atom laser.}\label{sec:CurrHigh}

In this section we consider the current and the outcoupled wave
function from excited modes in a harmonic trap. In particular we
consider a fermionic gas at $T=0$, but the formalism presented here
can also be used to analyze the contribution to the current and outcoupled
beam of the higher modes in a trapped BEC.

Quantum degeneracy was demonstrated for a trapped cloud of fermionic
alkali atoms in \cite{DeMarco1999a}. A gas of ultracold fermionic
atoms becomes superfluid \cite{Regal2004a,
Zwierlein2005a,Bourdel2004a,Partridge2005a,Chin2004a} by tuning the
interaction strength between two different hyperfine states.
Fermionic superfluidity relies on the formation of pairs of
attractive atoms. In the limit of weak interactions the system can
be described by the Bardeen-Cooper-Schrieffer (BCS) theory developed
in superconductivity that relates the order parameter to the binding
energy of the paired atoms (gap). A beam of atoms coherently
outcoupled from a trapped gas preserves the properties of the
initial state of the atoms. We explore the effect of quantum
degeneracy and fermionic superfluidity in the outcoupled beam
density profile and the total current. For the sake of simplicity we
consider  quasi one-dimensional systems. Furthermore, it is
reasonable to expect that the effect of superfluidity in the
outcoupling increases when the superfluid gap is highest along the
direction of gravity.

\subsection{Excited modes falling in the gravitational field}

In this section we present a one dimensional calculation of an
excited state of a harmonic trap falling in the gravitational field.
The current is calculated from the Franck-Condon factors of the
initial and final eigenfunctions. The initial Hamiltonian of a 1D
harmonic oscillator reads
\begin{equation}
H_{\rm trap}^{1D}=-\frac{\hbar^2}{2m}\frac{\rmd^2}{\rmd
z^2}+\frac{1}{2}m\omega_z^2z^2,
\end{equation}
whose eigenstates and eigenenergies are given by
\begin{eqnarray}
\psi_n(z)&=&\frac{1}{\pi^{1/4}\sqrt{a 2^n n!}}
\Her_n\left(\frac{z}{a}\right)\exp\left(-\frac{z^2}{2a_z^2}\right)
\label{psin}\\
E_n&=&\hbar \omega_z (n+\frac{1}{2}),
\end{eqnarray}
where $a=\sqrt{h/m\omega_z}$. The final state Hamiltonian is given
by the 1D version of $H_{\rm grav}$ defined in Eq.~(\ref{eq:Hcont})
\begin{equation}
H_{\rm f}= -\frac{\hbar^2}{2m}\frac{\rmd^2}{\rmd z^2}-Fz.
\end{equation}
The Hamiltonian $H_{\rm f}$ has a continuous spectrum
and the eigenfunctions can be labelled by the energy \cite{Lerme1990a}
\begin{equation} \label{eq:airy}
\psi_{E}^{\rm f}(z)=\langle z |\psi_E^{\rm f} \rangle = 2 \sqrt{F}
\beta \Ai\left(-2\beta F (z+\frac{E}{F})\right).
\end{equation}
Inserting the complete set of continuum eigenfunctions in
eq.~(\ref{eq:jtot_trace}) yields the following expression for the
total current originating from an initial state $\ket{\psi_n}$
\begin{equation}\label{eq:current1d}
J_n^{1D}(E)=\frac{2\pi\gamma^2}{\hbar} \int \rmd E_{\rm grav}
\delta(E-E_{\rm grav})|\langle \psi_n|\psi_E^{\rm f}\rangle|^2.
\end{equation}
The needed overlap integrals (which are Franck Condon factors) are
conveniently calculated by adapting a recursive method developed in
Ref.~\cite{Lerme1990a} (see also App.~\ref{app:A}):
\begin{align}
\langle \psi_{n}|\psi_E^{\rm f} \rangle&=& \sqrt{\frac{8
\sqrt{\pi}\beta \alpha}{2^n n!}} \exp{\left(\frac{16 \alpha^6}{3}-4
\beta E \alpha^2
\right)}\times\nonumber\\
&&K(n,2,-8\alpha^3,-2\beta E + 4 \alpha^4,-4 \alpha).
\end{align}
Here, $K(n,\alpha_L,\alpha'_L,\gamma_L,\delta_L)$ is given by eq.~(\ref{eq:K}).
The 1D current becomes
\begin{equation}\label{eq:CurrentTotal1Dz}
J_n^{1D}(E)=\frac{2\pi\gamma^2}{\hbar} |\langle
\psi_{n}|\psi_{E}^{\rm f}\rangle|^2.
\end{equation}
The outcoupled wave function can be calculated from
eq.~(\ref{eq:PsiContGreen})
\begin{equation}\label{eq:func1d}
\psi_{\rm out}^{1D}(n,z;E)= \gamma \int\rmd z'\; G_{1D}(z,z';E)
\psi_n(z').
\end{equation}
As shown in the appendix, a closed expression is given by
eq.~(\ref{eq:psin1d})
\begin{eqnarray}\label{eq:func1df}
 \psi_{\rm out}^{1D}(n,z;E)&=&B(n,z,E)\times\\
 && K(n,2,-8\alpha^3,-2\beta E + 4\alpha^4,-4\alpha),\nonumber
\end{eqnarray}
where $B(n,z,E)$ is defined in App.~\ref{app:A}.
\begin{figure}[t]
\includegraphics[width=\columnwidth]{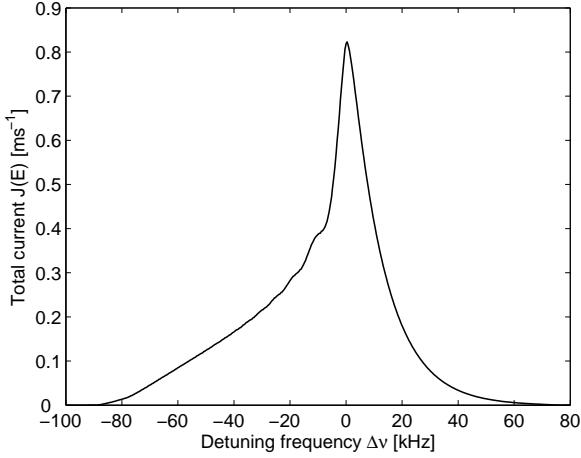}
\caption{Total 3D current for a quasi-1D Fermi gas
(eq.~(\ref{eq:Fcurr})) as a function of the detuning frequency
$\Delta\nu=E/h$. Parameters: $N=70$ spin-polarized atoms,
$\omega_z=2\pi 1.2$~kHz, $\omega_{\perp}=2 \pi 120$~kHz,
$E_F/h=84$~kHz, $T=0$, $\gamma/h = 100$~Hz,
$m_{\text{K}}=40$~u.\label{fig:jtot_fermi}}
\end{figure}
\subsection{Long axis in the direction of gravity}

We consider the current and outcoupled function for an initial state
of the form
\begin{equation}\label{eq:ini_zn}
\psi_{\rm ini}(x,y,z)=\langle \mbfr |\psi_{\rm ini} \rangle
=\psi_0(x)\psi_0(y)\psi_n(z)
\end{equation}
where $\psi_i$ are the eigenstates of the harmonic oscillator. Thus
we consider an arbitrary excited state along the $z$ direction while
the wavefunctions are gaussians in the perpendicular plane. The
total current is given by a convolution integral with the 1D current
of eq.~(\ref{eq:CurrentTotal1Dz}):
\begin{equation} \label{eq:tot_current}
J_n^{3D}(E) = 2 a_\perp^2\int_0^\infty\rmd
k_\perp\,k_\perp\rme^{-a_\perp^2 k_\perp^2}
J_n^{1D}\left(E-\frac{\hbar^2 k_\perp^2}{2m}\right).
\end{equation}
The corresponding 3D outcoupled wave-function is obtained by using
Eq.~(\ref{eq:psi3d})
\begin{eqnarray} \label{eq:fF3D}
\psi_{\rm out}^{3D}(\mbfr;E)
&=&\frac{a_\perp}{\sqrt{\pi}}\int_0^\infty\rmd k_\perp\,
k_\perp\rme^{-\frac{a_\perp^2 k_\perp^2}{2}}
\BesselJ_0(k_\perp\sqrt{x^2+y^2})
\times\nonumber\\
&&\psi_{\rm out}^{1D}(n,z;E-\frac{\hbar^2 k_\perp^2}{2m}).
\end{eqnarray}

\subsection{Fermi gas}

The wave function of a Fermi gas at $T=0$ is a Slater determinant of
the product state of the wave functions of $N$ atoms. The density
profile and the current are just a sum of the currents and wave
functions of the individual states of the atoms
\begin{equation}\label{eq:Fdens}
n_{\rm out}(\mbfr;E)=\sum_{n}f(E_n)|\psi_{\rm out}(n,\mbfr;E+E_n)|^2
\end{equation}
\begin{equation}\label{eq:Fcurr}
J(E)=\sum_{n}f(E_n)J_n(E+E_n),
\end{equation}
where $f(E_n)$ is the Fermi distribution function. For a BCS
superfluid gas with finite gap $\Delta$ the outcoupled density and
current for each spin can be calculated using the BCS distribution
function \cite{Tinkham1996b} $f_{\rm
BCS}(E_n)=1-{\xi_n}/{\sqrt{\xi_n^2+\Delta^2}}\tanh \left({
\sqrt{\xi_n^2+\Delta^2}}/{2 k_B T}\right)$ where $\xi_n=E_n-E_F$,
$E_F$ is the Fermi energy and $T$ the temperature of the system.

The current is shown in Fig.~\ref{fig:jtot_fermi} for a
spin-polarized normal gas. The 1D current in (\ref{eq:tot_current})
has an energy spread on the order of the Fermi energy
$E_F=N\hbar\omega_z$ due to the energy shifts of the contributions
of different modes. The convolution with the transverse direction
results in an even wider spread of the total current, which is caused
by the tight confinement in the transverse direction, as explained
in Sec.~\ref{sec:GeometryEffects}.  It was predicted in
\cite{Torma2000a} and demonstrated experimentally
\cite{Chin2004a,Kinnunen2004a} that superfluidity in a Fermi gas
leads to an energy shift in the outcoupled current when a RF (or
laser) field transfers atoms from one of the paired hyperfine states
to another hyperfine state. The shift originates in the additional energy
needed to outcouple the atoms that are forming Cooper pairs. In the
BCS regime considered here the gap cannot exceed a small percentage
of the Fermi energy. Such values would not create an appreciable
shift in the current in Fig~\ref{fig:jtot_fermi}. However, in a
strongly interacting fermionic superfluid \cite{Regal2004a,
Zwierlein2005a,Bourdel2004a,Partridge2005a} the gap can reach values
of the order of $0.2 E_F$ \cite{Chin2004a,Kinnunen2004a} that would
lead to an experimentally detectable shift in the current in
Fig.~\ref{fig:jtot_fermi}.\\
\begin{figure}[t]
\includegraphics[width=\columnwidth]{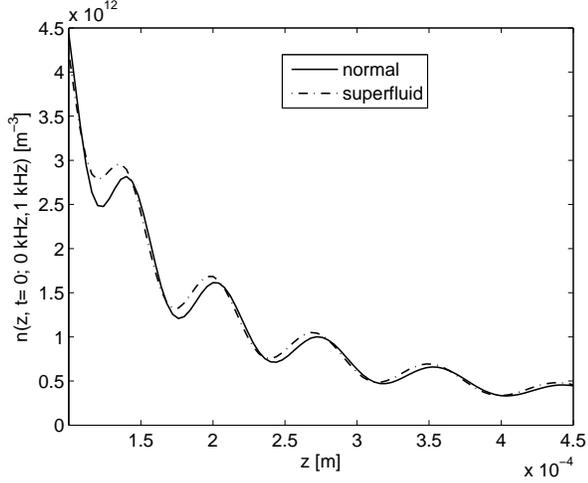}
\caption{ Density profile of the fermionic beam outcoupled with two
different RF $|E_{\rm grav}-E_{\rm trap}|/h$ and 1kHz+$(E_{\rm
grav}+E_{\rm trap})/h$. We show the density profile
(eq.~(\ref{eq:corr})) at the center of the beam for a Fermi gas with
$N=70$, $\omega_z= 2\pi 1.2$~kHz, $\omega_{\perp}=2 \pi 120$~kHz,
and pairing gap $\Delta=3.36$~kHz, $\gamma/h=100$Hz,
$m_{\text{K}}=40$~u.\label{fig:corr}}
\end{figure}
The one particle correlation of the BEC was measured experimentally
by outcoupling particles from the BEC with two different RF
\cite{Bloch2000a}. An equivalent process in a Fermi gas would lead
to a density profile of the form
\begin{eqnarray}\label{eq:corr}
{\rm n}(\mbfr,t;\nu_1,\nu_2)&=&\sum_{n}f(E_n)|\psi_{\rm
out}(n,\mbfr;h \nu_1+E_n)e^{i h \nu_1 t} \nonumber\\&& + \psi_{\rm
out}(n,\mbfr;h\nu_2+E_n)e^{i h \nu_2 t} |^2.
\end{eqnarray}
As demonstrated in Fig.~\ref{fig:corr}, the density profile at the center of
the beam shows oscillations. For a Fermi gas, the oscillations are
not related to superfluidity as in a BEC (see (\ref{eq:twobeams}) and \cite{Bloch2000a}). In 1D,
the density profile (\ref{eq:corr}) is a sum of oscillatory
functions in $E+Fz$ (see eq.~(\ref{eq:func1d})). Each of the terms
in the sum is shifted by $\hbar \omega_z$, but because $E_F \ll Fz$
the period of oscillation is effectively the same. The 3D density
profile (\ref{eq:fF3D}) is a convolution of the 1D density
function with the energy of the transverse directions that would
result in oscillations on the scale $E+Fz$.
Each term of the
sum will contribute again by shifting the rescaled energy and as
long as $E_F \ll Fz$ all of them will have effectively the same
period of oscillation. As expected, the one particle correlation
function does not show any effect of the superfluidity. The gap
energy shift transforms into a shift $\delta z=\Delta/F$ (see
Fig.~\ref{fig:corr}) that only leads to a time shift in the density
profiles. Fermionic superfluidity relies on the formation of atomic
pairs and therefore one expects to see some effect of superfluidity
only in the two particle correlation function \cite{Greiner2005a}.

\section{Summary and outlook}

A quantum mechanical theory of the atom laser based on
propagator techniques has been presented and contrasted with various
approximation schemes. The analysis of the current and outcoupled
beam leads to a distinction between small and big condensates. The
experimentally observed structure and substructure
\cite{Koehl2005a,Riou2005a} of the transverse beam profile has been
obtained using the T-matrix formalism, which includes the effect of
interactions.

The effect of a non-isotropic geometry of the trap has been analyzed
and a simple way to convolute the propagation along the
gravitational and transverse direction has been determined. We have
extended the formalism to calculate the current and outcoupled beam
from excited modes. We have applied it to calculate the current and
beam profile for a quasi-1D Fermi gas. We have shown that the
interference pattern of atoms outcoupled with two different RF show
oscillations that are not related to superfluidity as in the case of
a Bose gas. The effect of fermionic superfluidity in the current and
outcoupled beam has been discussed.

\begin{appendix}

\section{Recursion relations for the current and beam density from the excited modes of a trap}\label{app:A}

We derive eq.~(\ref{eq:func1df}) for the outcoupled wavefunction
from an excited state in a one dimensional harmonic trap. The
generating function of the Hermite polynomials reads
\begin{equation}
\Her_n\left(\frac{z}{a}\right)=\left[\frac{\partial^n}{\partial
t^n}\exp(-t^2+2tz/a)\right]_{t=0}.
\end{equation}
Inserting it into Eq.(\ref{eq:func1d}) leads to
\begin{eqnarray}
&&\psi_{\rm out}^{1D}(n,z,E)
= \gamma \int \rmd z' G_{\rm grav}^{1D}(z,z',E)\psi_n(z')\\
&&=\gamma \frac{1}{\sqrt{2^n n! a \sqrt{\pi}}}\int \rmd z' G(z,z',E)
\Her_n(\frac{z'}{a})\rme^{-\frac{z'^2}{2a^2}}\nonumber\\
&&= \gamma \frac{1}{\sqrt{2^n n!}} \left[ \frac{\partial^n}{\partial
t^n} \rme^{-t^2}  \frac{\int \rmd z' G(z,z',E)\rme^{-\frac{z'^2}{2
a^2}+2t\frac{z'}{a}}}{\sqrt{a}\pi^{1/4}} \right]_{t=0}. \nonumber
\end{eqnarray}
Introducing a new variable $z''=z'+2ta$ and using the translation
law for the Green function \cite{Bracher2003a}, Eq.~(33), yields
\begin{eqnarray}
&&\psi_{\rm out}^{1D}(n,z,E)\\
&&\!\!\!\!\!\!\!\!\!\!=\frac{\gamma}{\sqrt{2^n n!}}
\left[\frac{\partial^n}{\partial t^n}\rme^{-t^2+2t^2} \frac{\int
\rmd z'' G(z,z''+2ta,E)\rme^{-\frac{z''^2}{2
a^2}}}{\sqrt{a}\pi^{1/4}}\right]_{t=0}\nonumber \\
&&\!\!\!\!\!\!\!\!\!\!=\frac{\gamma}{\sqrt{2^n n!}}
\left[\frac{\partial^n}{\partial t^n}\rme^{t^2} \frac{\int \rmd
z''G(z-2ta,z'',E+2taF)\rme^{-\frac{z''^2}{2
a^2}}}{\sqrt{a}\pi^{1/4}}\right]_{t=0}\nonumber
\end{eqnarray}
The $z''$ integration in the last expression yields the known
outgoing wave function for $n=0$,
\begin{eqnarray}\label{eq:psin1d}
&&\psi_{\rm out}^{1D}(n,z,E)\nonumber\\
&&=\frac{\gamma}{\sqrt{2^n n!}} \left[\frac{\partial^n}{\partial
t^n}\rme^{t^2}
\psi_{\rm out}^{1D}(0,z-2ta,E+2ta)\right]_{t=0}\nonumber\\
&&=-\frac{\gamma}{\sqrt{2^n n!}}4\sqrt{2} \pi^{5/4} \sqrt{F\alpha
\beta^3}
\rme^{-8\alpha^6/3+2\alpha^2\tilde{\epsilon}}\Ci(\tilde{\epsilon}-2\tilde{\xi}
)\times\nonumber\\
&&\qquad\left[\frac{\partial^n}{\partial t^n}\rme^{t^2-8\alpha^3t}
\Ai(\tilde{\epsilon}-4\alpha t) \right]_{t=0}\nonumber\\
&&=B(n,z,E) \left[ K(n,2,-8\alpha^3,-2\beta E + 4\alpha^4,-4\alpha)
\right].
\end{eqnarray}
Here, we separated the factors in front of the square brackets from
the derivatives and used the $K(\ldots)$ notation of
Ref.~\cite{Lerme1990a}, eq.~(13):
\begin{equation}\label{eq:K}
K(n,\alpha_L,\alpha'_L,\gamma_L,\delta_L)
=\left[\frac{\partial^n}{\partial t^n} \rme^{\frac{1}{2}\alpha_L
t^2+\alpha'_L t} \Ai(\gamma_L + \delta_L t)\right]_{t=0}
\end{equation}
The expressions $K(n)$ are readily calculated using a recursion
relation derived in \cite{Lerme1990a} that we show here for
completeness. One can define
\begin{equation}
K'(n,\alpha_L,\alpha'_L,\gamma_L,\delta_L)=
\left[\frac{\partial^n}{\partial t^n}\rme^{\frac{1}{2}\alpha_L
t^2+\alpha'_Lt} \Ai'(\gamma_L + \delta_L t)\right]_{t=0},
\end{equation}
and simultaneously calculate $K(n)$ and $K'(n)$:
\begin{eqnarray}\label{eq:recursion}
K(0)&=&\Ai(\gamma_L), \quad K(1)=\alpha'_L K(0)+\delta_L K'_L (0), \nonumber \\
K'(0)&=&\Ai'(\gamma_L), \quad K'(1)=\alpha'_L K'(0)+\delta_L\gamma_L
K(0),
\nonumber \\
K(n)&=&\alpha_L'K(n-1)+\alpha_L (n-1)K(n-2)\nonumber\\
   &&+\delta_L K'(n-1), \nonumber \\
K'(n)&=&\alpha'_LK'(n-1)+\alpha_L(n-1)K'(n-2)\nonumber\\
    &&+\delta_L\gamma_LK(n-1)+\delta_L^2(n-1)K(n-2).
\end{eqnarray}
As pointed out in \cite{Lerme1990a} the recursion method is unstable
for $|\delta_L|<1/2$ and for large $\gamma_L$. This means that we
cannot use this method to calculate the current and outcoupled beam
profile for small trapping frequencies or large number of atoms. For
small trapping frequencies, one could use the reflection
approximation eq.~(\ref{eq:jtot_ideal_refl}).
\end{appendix}

\section*{Acknowledgements}

We appreciate helpful discussions with C.~Bracher, M.~Kleber, and P.~Kramer. This work is supported by the Deutsche Forschungsgemeinschaft (grant KR~2889 [Emmy Noether Program]), by the EPSRC through project EP/C51933/1 and the ESF (BEC2000+). TK would like to thank M.~Moshinky for the invitation and hospitality at UNAM during the completion of this work.

\providecommand{\url}[1]{#1}

\end{document}